\documentclass[12pt]{article}
\usepackage{graphicx}

\begin{document}
\begin{center}
{\Large\bf Productions of $J/\psi$ mesons in $p$-Pb collisions at
5 TeV}

\vskip1.0cm

Fu-Hu Liu$^{a,}${\footnote{E-mail: fuhuliu@163.com;
fuhuliu@sxu.edu.cn}}, Hai-Ling Lao$^{a}$, and Roy A.
Lacey$^{b,}${\footnote{E-mail: Roy.Lacey@Stonybrook.edu}}

{\small\it $^a$Institute of Theoretical Physics, Shanxi
University, Taiyuan, Shanxi 030006, China

$^b$Departments of Chemistry \& Physics, Stony Brook University,
Stony Brook, NY 11794, USA}
\end{center}

\vskip1.0cm

{\bf Abstract:} The rapidity distributions of $J/\psi$ mesons
produced in proton-lead ($p$-Pb) collisions at center-of-mass
energy per nucleon pair $\sqrt{s_{NN}}=5$ TeV are studied by using
a multisource thermal model and compared with the experimental
data of the LHCb and ALICE Collaborations. Correspondingly, the
pseudorapidity distributions are accurately obtained from the
parameters extracted from the rapidity distributions. At the same
time, the transverse momentum distributions in the same
experiments are described by the simplest Erlang distribution
which is the folding result of two exponential distributions which
are contributed by the target and projectile partons respectively.
\\
\\

{\section{Introduction}}

The successful running of the Large Hadron Collider (LHC) at the
European Organization for Nuclear Research (CERN) has been
advancing heavy ion (nucleus-nucleus) collisions from GeV to TeV
energies [1--4]. It is already established that a new state of
matter, namely the Quark-Gluon Plasma (QGP), has been formed in
nucleus-nucleus collisions at high energies due to high
temperature and density [5--11]. As one of the most valuable
signatures of the formation of QGP, the suppression of $J/\psi$
mesons produced in nucleus-nucleus collisions has been studied
[12--15]. In fact, the suppression of $J/\psi$ mesons can also be
found in proton-nucleus ($pA$) collisions, where the QGP is not
expected to produce [16] due to small system. Instead, some cold
nuclear matter effects such as nuclear absorption and shadowing as
well as parton energy loss affect the productions of final-state
particles in $pA$ collisions [17--19]. In addition, at LHC
energies, it is well established by now that the situation is much
more involved, with recombination processes playing an essential
role [20, 21].

Except for the production of QGP in high energy nucleus-nucleus
collisions, other topics such as some universal laws and
particular properties of measurable quantities in intermediate and
high energy particle-particle, particle-nucleus, and
nucleus-nucleus collisions are interested for the community of
particle and nuclear physics. These universal laws and particular
properties are expected to exist in transverse momentum
distribution, multiplicity and transverse energy distribution,
rapidity distribution and correlation, azimuthal distribution and
correlation, fragment production, and so forth [22--26]. We have
always been interested in the study of universal laws in different
systems [22--24].

Recently, the productions of $J/\psi$ mesons in proton-lead
($p$-Pb) collisions at LHC energies are studied. Some experimental
results are obtained [16, 27--30]. We are interested in the
rapidity ($y$) and transverse momentum ($p_T$) distributions of
$J/\psi$ mesons. From $y$ distribution, we can test some models in
the longitudinal rapidity space, and obtain some information on
energy loss of partons and penetrating (stopping) power of
projectile and target nuclei. From $p_T$ distribution, we can test
some models in the transverse momentum space and obtain excitation
degree of the interacting system.

In this paper, by using a multisource thermal model [31--33], we
study $y$ and $p_T$ distributions of $J/\psi$ mesons produced in
$p$-Pb collisions at center-of-mass energy per nucleon pair
$\sqrt{s_{NN}}=5$ TeV which is one of the LHC energies
corresponding to a proton beam energy of 4 TeV and a lead beam
energy of 1.58 TeV per nucleon. At the same time, the
pseudorapidity ($\eta$) distributions of $J/\psi$ mesons are
obtained. In section 2, a description of the model and calculation
method is presented. In section 3, the results and discussion are
given. The calculated results are found to be in agreement with
the available experimental data of the LHCb and ALICE
Collaborations [16, 29, 30]. Finally, we summarize our main
observations and conclusions in section 4.
\\

{\section{The model and calculation method}}

The model employed in the present work is the multisource thermal
model [31--33] which is a successor of the thermalized cylinder
model [34, 35] which is based on the one-dimensional string model
[36] and the fireball model [37]. According to the one-dimensional
string model [36], in high energy nucleon-nucleon collisions, a
string is formed consisting of two endpoints acting as energy
reservoirs and the interior with constant energy per length.
Because of the asymmetry of the mechanism, the string will break
into many substrings along the direction of incident beam.
According to the fireball model [37], in the mentioned collisions,
the incident nucleon penetrates through the target nucleon, then a
fire streak (a series of fireballs) is formed along the direction
of incident beam. The distribution length of substrings in the
one-dimensional string model [36] and the length of fire streak in
the fireball model [37] will define the width of the
(pseudo)rapidity distribution. In high energy nucleus-nucleus
collisions, many strings or fire streaks are formed along the
incident direction. Finally, a thermalized cylinder is formed
because of these strings or fire streaks mix in the transverse
direction.

Due to different excitation degrees of substrings or fireballs,
the interacting system which contains many substrings or fireballs
can be divided into several regions or sources. In addition,
different interacting mechanisms or event samples can be resulted
in different sources. Each source contains several sub-sources
which can be substrings, fireballs, partons, or nucleons due to
different topics of investigations such as the distributions of
transverse momenta, multiplicities, rapidities, transverse
energies, etc. Different sources can be described by the same law
with different parameters or by different laws. The distribution
in final state is usually contributed by the several sources,
which results in a multi-component distribution which results from
the multisource thermal model.

In the framework of the considered model, most of light flavor
particles such as pions and kaons can be regarded as a result of
soft excitation process due to thermal reason. As heavy quark
particle, $J/\psi$ is produced inherently in a hard process which
proceeds through parton-parton collisions. We assume that a parton
in target nucleus (nucleon) and a parton in projectile nucleus
(nucleon) take part in the collisions to form the source to emit
$J/\psi$ meson. Many sources can be formed in nucleus-nucleus
collisions and in the considered data sample. These sources can
appear in different regions in the interacting overlapping area.
In rapidity space, in the laboratory or center-of-mass reference
frame, these sources distribute at different rapidities ($y_x$)
due to different rapidity shifts.

The sources with $y_x<0$ are in the backward region which are
mainly contributed by the target nucleus, and the sources with
$y_x>0$ are in the forward region which are mainly contributed by
the projectile nucleus. The backward and forward regions are
expected in $[y_T,0]$ and $[0,y_P]$ respectively, where $y_T$
(which is less than 0) and $y_P$ (which is larger than 0) denote
the maximum rapidity shifts in the backward and forward regions
respectively, i.e. $y_T$ is the minimum $y_x$ and $y_P$ is the
maximum $y_x$. We would like to point out that the separation for
the backward and forward regions does not mean that there is no
source in the mid-rapidity region. In fact, these sources can also
be divided into three groups: a central region with sources around
the mid-rapidity, a target fragmentation region with sources in
the target side, and a projectile fragmentation region with
sources in the projectile side. The sources in the same region
form a large source. Then, we have a three-source picture which is
compatible with previous works [38--49].

Each parton (the $i$-th parton) is assumed to contribute an
exponential transverse momentum ($p_{Ti}$) distribution with a
mean value of $\langle p_{Ti} \rangle$. The mentioned distribution
is
\begin{equation}
f_i(p_{Ti})=\frac{1}{\langle p_{Ti} \rangle} \exp \biggr(
-\frac{p_{Ti}}{\langle p_{Ti} \rangle \large} \biggr),
\end{equation}
where $i=1$ and 2 for the target parton and projectile parton
respectively. Generally, $\langle p_{T1} \rangle = \langle p_{T2}
\rangle = \langle p_{Ti} \rangle$. The $p_T$ $(=p_{T1}+p_{T2})$
distribution of $J/\psi$ is the folding result of two exponential
distributions. We have $p_T$ distribution to be the simplest
Erlang distribution
\begin{equation}
f(p_{T})= \int_0^{p_T} f_1(p_{T1})f_2(p_T-p_{T1}) dp_{T1}
=\int_0^{p_T} \frac{1}{\langle p_{Ti} \rangle ^2} \exp \biggr(
-\frac{p_{T}}{\langle p_{Ti} \rangle} \biggr) dp_{T1} =
\frac{p_T}{\langle p_{Ti} \rangle ^2} \exp \biggr(
-\frac{p_{T}}{\langle p_{Ti} \rangle} \biggr).
\end{equation}

In the Monte Carlo method, according to $\int_0^{p_{T1,2}}
f_{1,2}(p_{T1,2})dp_{T1,2} = R_{1,2}$, we have $p_{T1,2}= -
\langle p_{Ti} \rangle \ln (1-R_{1,2})$, where $R_{1,2}$ denote
random numbers in [0,1]. Because of both $1-R_{1,2}$ and $R_{1,2}$
being random numbers in [0,1], we have
\begin{equation}
p_T=-\langle p_{Ti} \rangle (\ln R_1 +\ln R_2).
\end{equation}
As a statistical result, in the source rest frame, we assume that
$J/\psi$ mesons are isotropically emitted, which results in the
distribution of polar angle $\theta'$ being $\frac{1}{2}\sin
\theta'$. Then, the polar angle $\theta'$ satisfies
$\int_0^{\theta'} \frac{1}{2}\sin \theta' d\theta' =R_3$ in the
Monte Carlo method, where $R_3$ denotes random numbers in [0,1].
We have $\theta'$ to be
\begin{equation}
\theta'=\arctan \Biggl[ \frac{2\sqrt{R_3(1-R_3)}}{1-2R_3}
\Biggr]+\theta_0,
\end{equation}
where $\theta_0=0$ (or $\pi$) is for the case of the first term
being larger than 0 (or less than 0) in Eq. (4). The longitudinal
momentum $p'_z$ and energy $E'$ in the rest frame can be expressed
as
\begin{equation}
p'_z=p_T \cot \theta'
\end{equation}
and
\begin{equation}
E'=\sqrt{p_T^2+p'^2_z+m_0^2}
\end{equation}
respectively, where $m_0$ denotes the rest mass of the considered
particle.

In the laboratory or center-of-mass reference frame, the rapidity
$y$, longitudinal momentum $p_z$, polar angle $\theta$, and
pseudorapidity $\eta$ of the considered particle can be given by
\begin{equation}
y=\frac{1}{2} \ln \Biggr( \frac{E'+p'_z}{E'-p'_z} \Biggr) +y_x,
\end{equation}
\begin{equation}
p_z=\sqrt{p_T^2+m_0^2}\sinh y,
\end{equation}
\begin{equation}
\theta =\arctan(p_T/p_z),
\end{equation}
and
\begin{equation}
\eta=-\ln \tan(\theta/2),
\end{equation}
respectively. The rapidity, pseudorapidity, and transverse
momentum distributions are then given by the statistical method.
In particular, for rapidity (pseudorapidity) distribution, the
contribution fraction (relative contribution) $k_T$ of the
backward region and the contribution fraction $1-k_T$ of the
forward region may be different due to asymmetric $p$-Pb
collisions.
\\

{\section{Results and discussion}}

Fig. 1(a) presents the rapidity distributions, $d\sigma/dy$, of
$J/\psi$ mesons produced directly from the proton-nucleon
collisions (prompt $J/\psi$) and from $b$-hadron decays ($J/\psi$
from $b$) in $p$-Pb collisions at 5 TeV, where $d\sigma$ denotes
the production cross-section of the considered $J/\psi$ in
rapidity bin $dy$. The symbols represent the experimental data of
the LHCb Collaboration [16] and the curves are our fitting results
based on the Monte Carlo calculation. In the calculations, for
both the process of $J/\psi$ productions, we take $\langle
p_{Ti}\rangle=1.50\pm0.10$ GeV/$c$; for the process of prompt
$J/\psi$, we take $y_T= -4.41\pm0.40$, $y_P= 3.78\pm0.40$, $k_T=
0.58\pm0.06$, and $\sigma_0=(5099.8\pm350.0)$ $\mu$b with $\chi^2$
per degree of freedom ($\chi^2$/dof) to be 1.394, where $\sigma_0$
denotes the total production cross-section of the considered
$J/\psi$ in full rapidity space; and for the process of $J/\psi$
from $b$, we take $y_T= -3.86\pm0.40$, $y_P= 3.67\pm0.35$, $k_T=
0.50\pm0.04$, and $\sigma_0=(629.2\pm41.0)$ $\mu$b with
$\chi^2$/dof to be 2.299. The normalization factor is in fact the
production cross-section in full rapidity range. One can see that
the model describes the experimental data of the LHCb
Collaboration.

\begin{figure}
\hskip-1.0cm \begin{center}
\includegraphics[width=16.0cm]{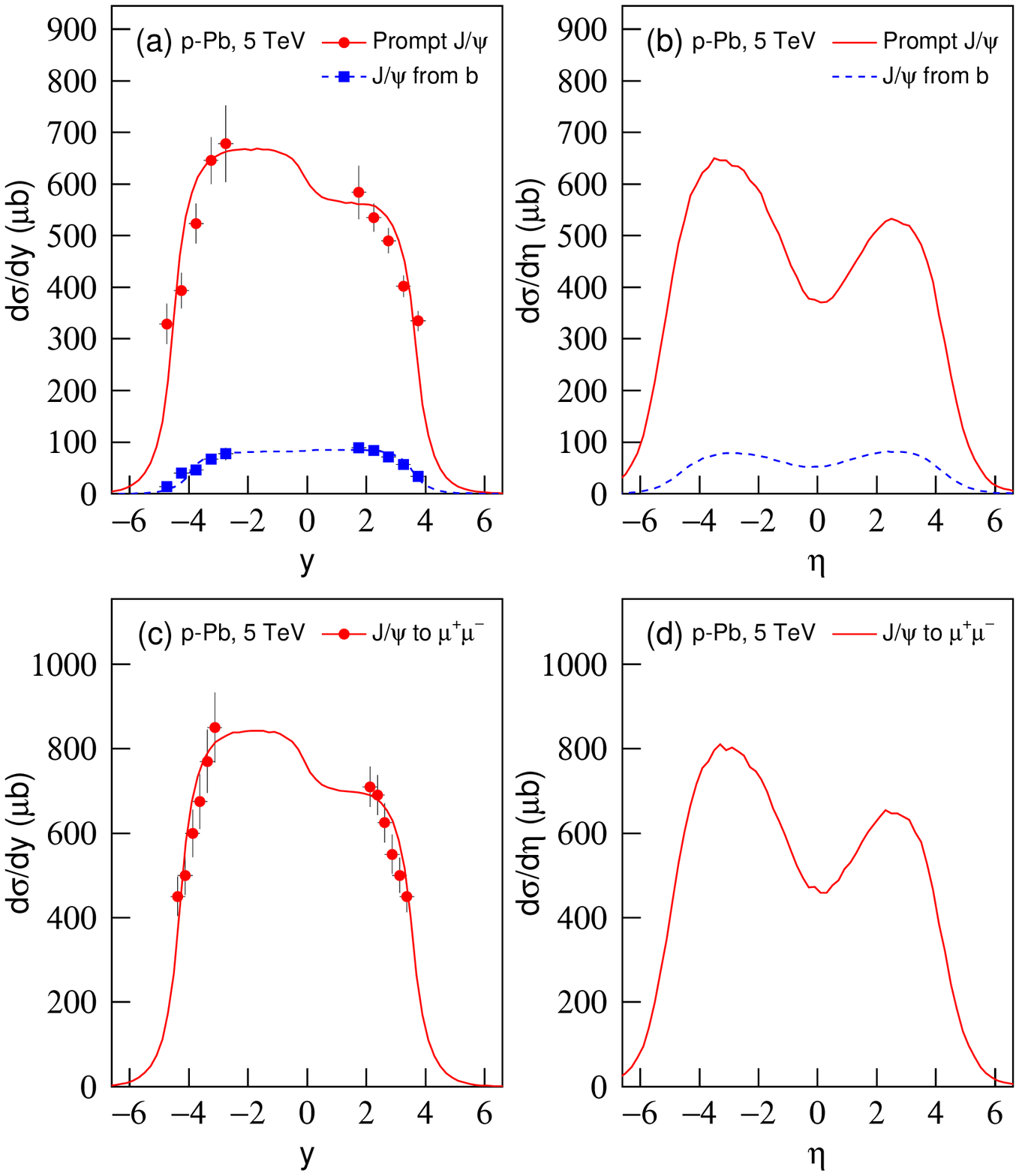}
\end{center}
\vskip0.0cm Fig. 1. (a) Rapidity distributions of prompt $J/\psi$
and $J/\psi$ from $b$ in $p$-Pb collisions at 5 TeV. The symbols
represent the experimental data of the LHCb Collaboration [16] and
the curves are our calculated results. (b) Pseudorapidity
distributions corresponding to the rapidity curves in Fig. 1(a).
(c) Rapidity distributions of inclusive $J/\psi$ to $\mu^+\mu^-$
in $p$-Pb collisions at 5 TeV. The symbols represent the
experimental data of the ALICE Collaboration [29] and the curve is
our calculated result. (d) Pseudorapidity distribution
corresponding to the rapidity curve in Fig. 1(c).
\end{figure}

To see the characteristics of pseudorapidity distributions, the
results corresponding to the curves in Fig. 1(a) are given in Fig.
1(b). Both the results for $y$ and $\eta$ distributions are direct
and accurate. There is no conversion between them, where an
unsuitable conversion may cause errors [52]. One can see large
hollow structure in the region of $\eta=0$. The difference between
$y$ and $\eta$ is obvious for the production of heavy particles
such as $J/\psi$. We cannot use $y\approx \eta$ in our calculation
for heavy particles even at LHC energies.

Figs. 1(c) and 1(d) are similar to Figs. 1(a) and 1(b)
respectively, but the former two are for inclusive $J/\psi$ to
$\mu^+\mu^-$ only. The symbols represent the experimental data of
the ALICE Collaboration [29] and the curves are our modelling
results. In the calculation, we take $\langle
p_{Ti}\rangle=1.40\pm0.08$ GeV/$c$, $y_T=-4.20\pm0.40$, $y_P=
3.67\pm0.40$, $k_T= 0.58\pm0.06$, and $\sigma_0=(6135.5\pm510.0)$
$\mu$b with $\chi^2$/dof to be 0.454. One can see that the model
describes the experimental data of the ALICE Collaboration. Again,
the difference between $y$ and $\eta$ is obvious for the
production of $J/\psi$.

From Figs. 1(a) and 1(c), one can see that the sources for the
creations of prompt $J/\psi$, $J/\psi$ from $b$, and inclusive
$J/\psi$ to $\mu^+\mu^-$ have nearly the same rapidity shift in
the uncertainty range. For each creation, the rapidity shift in
the backward region seems to be greater than that in the forward
one, though large uncertainty range is used. Our calculation based
on a revised nuclear-collision geometry [50] shows that the mean
number of $p$-nucleon collisions in $p$-Pb collisions is 2.7. If
the mean energy loss ratios in the first (or last) and other
$p$-nucleon collisions are 98.28\% and 52.20\% respectively, which
are about two times of those (50\% and 25\%) in fixed target
experiments [51], the energies of each participant nucleon after
collisions in the backward and forward regions are
1.58$\times$0.0172 TeV and 4$\times$0.0172$\times$0.4780$^{1.7}$
TeV, and the corresponding velocities $\beta$ are 0.99940$c$ and
0.99886$c$, respectively. Thus, the mean rapidity shifts
($y=0.5\ln[(1+\beta)/(1-\beta)]$) in the backward and forward
regions are $-4.06$ and 3.73 respectively, which are consistent to
$y_T$ and $y_P$ respectively used in the present work.

Figs. 2(a) and 2(b) show the transverse momentum distributions,
$d\sigma/dp_T$, of prompt $J/\psi$ (and $J/\psi$ from $b$) in
rapidity ranges $1.5<y<4.0$ and $-5.0<y<-2.5$ in $p$-Pb collisions
at 5 TeV respectively. The symbols represent the experimental data
of the LHCb Collaboration [16], the solid curves are our fitting
results based on Eq. (2), and the dashed curves will be discussed
later. The values of related parameters and $\chi^2$/dof for the
solid curves are presented in Table 1. One can see that the $p_T$
distributions obey the simplest Erlang distribution. The value of
$\langle p_{Ti} \rangle$ for prompt $J/\psi$ is less than that for
$J/\psi$ from $b$, where the later one needs larger threshold
energy for creation of $b$-hadron.

Figs. 2(c) and 2(d) are similar to Fig. 2(a), but the former two
are for inclusive $J/\psi$ to $\mu^+\mu^-$ and $e^+e^-$
respectively, measured by the ALICE Collaboration [30] in
different rapidity ranges shown in the panel and with alternative
expression ($d^2\sigma/dydp_T$) of transverse momentum
distribution. Particularly, in Fig. 2(d), only the solid curve
based on Eq. (2) for $J/\psi$ to $e^+e^-$ is presented. The values
of related parameters and $\chi^2$/dof are listed in Table 1. Once
again, the $p_T$ distributions obey the simplest Erlang
distribution. The value of $\langle p_{Ti} \rangle$ in large $|y|$
region is less than that in small $|y|$ region, where large angle
scattering appears in small $|y|$ region which results in large
$p_T$.

To see clearly the dependence of the transverse momentum
distribution on rapidity, Figs. 3(a) and 3(b) present
$d\sigma/dp_T$ versus $p_T$ for prompt $J/\psi$ and for $J/\psi$
from $b$ respectively, in different rapidity ranges. The symbols
represent the experimental data of the LHCb Collaboration [16],
the solid curves are our fitting results based on Eq. (2), and the
dashed curves will be discussed later. For the purpose of
clearness, the results for different rapidity ranges are
multiplied by different amounts as marked in the panels. The
values of related parameters and $\chi^2$/dof for the solid curves
are listed in Table 1. Once more, the $p_T$ distributions obey the
simplest Erlang distribution. The value of $\langle p_{Ti}
\rangle$ for prompt $J/\psi$ is less than that for $J/\psi$ from
$b$, and both the values of $\langle p_{Ti} \rangle$ decrease with
increase of the rapidity.
\\

{\small {Table 1. Values of parameters and $\chi^2$/dof
corresponding to the solid curves in Figs. 2 and 3.
{%
\begin{center}
\begin{tabular}{ccccc}
\hline\hline  Fig. & Type or $y$ range & $\langle p_{Ti}\rangle $ (GeV/$c$) & $\sigma_0$ ($\mu$b) & $\chi^2$/dof \\
\hline
2(a) & prompt $J/\psi$     & $1.38\pm0.08$ & $1163.8\pm120.4$ & 0.632 \\
     & $J/\psi$ from $b$   & $1.61\pm0.13$ & $165.2\pm17.8$   & 0.317 \\
2(b) & prompt $J/\psi$     & $1.24\pm0.05$ & $1300.6\pm138.1$ & 0.954 \\
     & $J/\psi$ from $b$   & $1.51\pm0.11$ & $114.0\pm12.6$   & 0.297 \\
2(c) & $2.03<y<3.53$       & $1.42\pm0.10$ & $583.9\pm60.3$   & 0.505 \\
     & $-4.46<y<-2.96$     & $1.25\pm0.05$ & $638.4\pm58.5$   & 0.529 \\
2(d) & $J/\psi$ to $e^+e^-$& $1.45\pm0.11$ & $933.5\pm90.2$   & 0.013 \\
3(a) & $1.5<y<2.0$   & $1.48\pm0.10$ & $585.4\pm60.0$   & 0.203 \\
     & $2.0<y<2.5$   & $1.49\pm0.10$ & $539.3\pm52.9$   & 0.308 \\
     & $2.5<y<3.0$   & $1.45\pm0.09$ & $488.0\pm45.1$   & 0.341 \\
     & $3.0<y<3.5$   & $1.33\pm0.07$ & $404.1\pm40.3$   & 0.559 \\
     & $3.5<y<4.0$   & $1.26\pm0.06$ & $337.8\pm23.5$   & 0.518 \\
3(b) & $1.5<y<2.0$   & $1.74\pm0.15$ & $89.2\pm9.9$     & 0.120 \\
     & $2.0<y<2.5$   & $1.69\pm0.14$ & $84.1\pm8.9$     & 0.395 \\
     & $2.5<y<3.0$   & $1.61\pm0.13$ & $70.9\pm6.7$     & 0.402 \\
     & $3.0<y<3.5$   & $1.49\pm0.10$ & $55.8\pm4.6$     & 0.520 \\
     & $3.5<y<4.0$   & $1.47\pm0.10$ & $35.3\pm3.4$     & 0.357 \\
\hline\hline
\end{tabular}%
\end{center}
}} }

\vskip0.5cm

\begin{figure}
\hskip-1.0cm \begin{center}
\includegraphics[width=16.0cm]{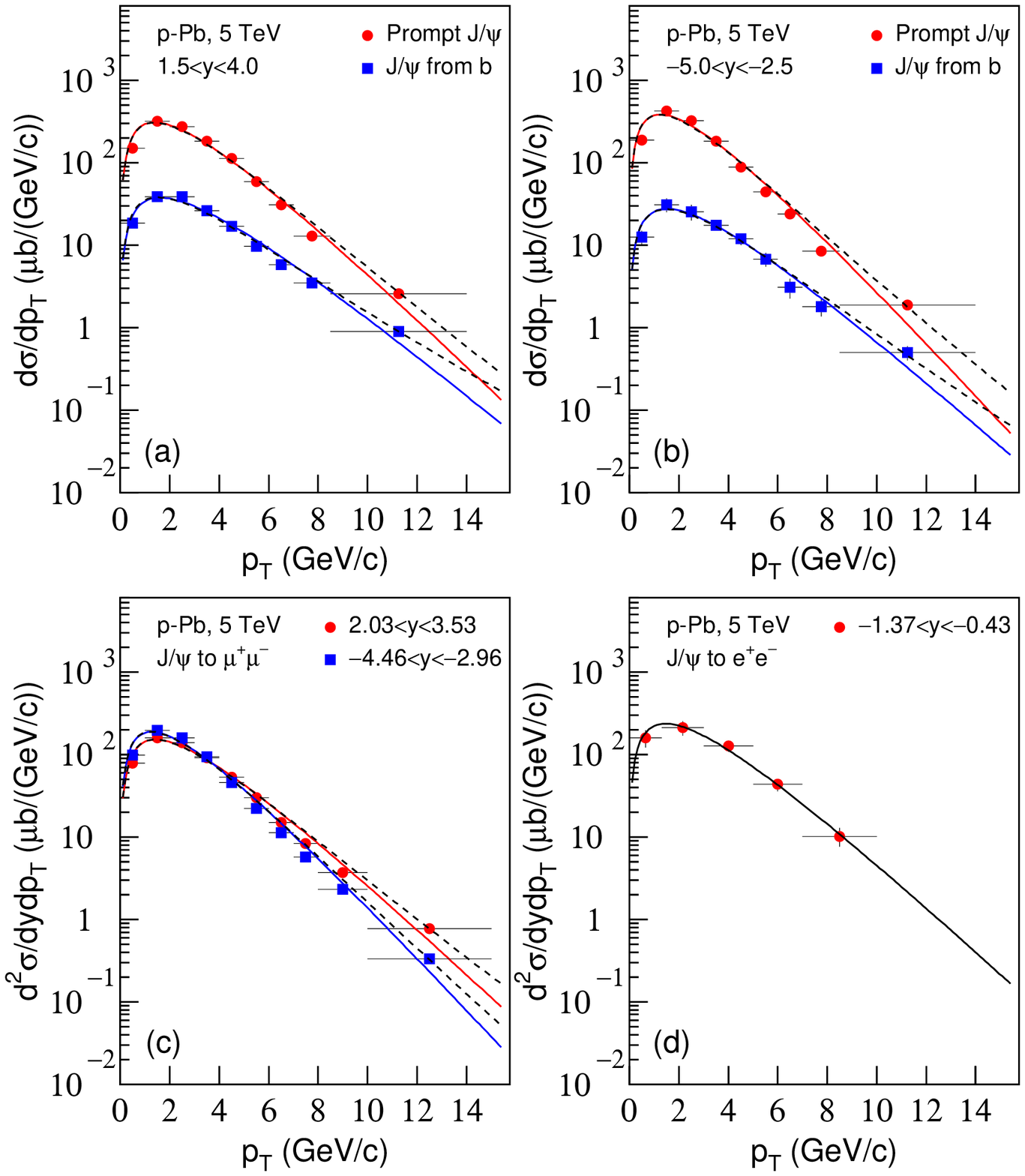}
\end{center}
\vskip0.0cm Fig. 2. (a)(b) Transverse momentum distributions of
prompt $J/\psi$ and $J/\psi$ from $b$ in rapidity ranges (a)
$1.5<y<4.0$ and (b) $-5.0<y<-2.5$ in $p$-Pb collisions at 5 TeV.
The symbols represent the experimental data of the LHCb
Collaboration [16] and the curves are our calculated results.
(c)(d) Transverse momentum distributions of inclusive $J/\psi$ to
(c) $\mu^+\mu^-$ and (d) $e^+e^-$ in $p$-Pb collisions at 5 TeV.
The symbols represent the experimental data of the ALICE
Collaboration [30] and the curves are our calculated results.
\end{figure}

\begin{figure}
\hskip-1.0cm \begin{center}
\includegraphics[width=16.0cm]{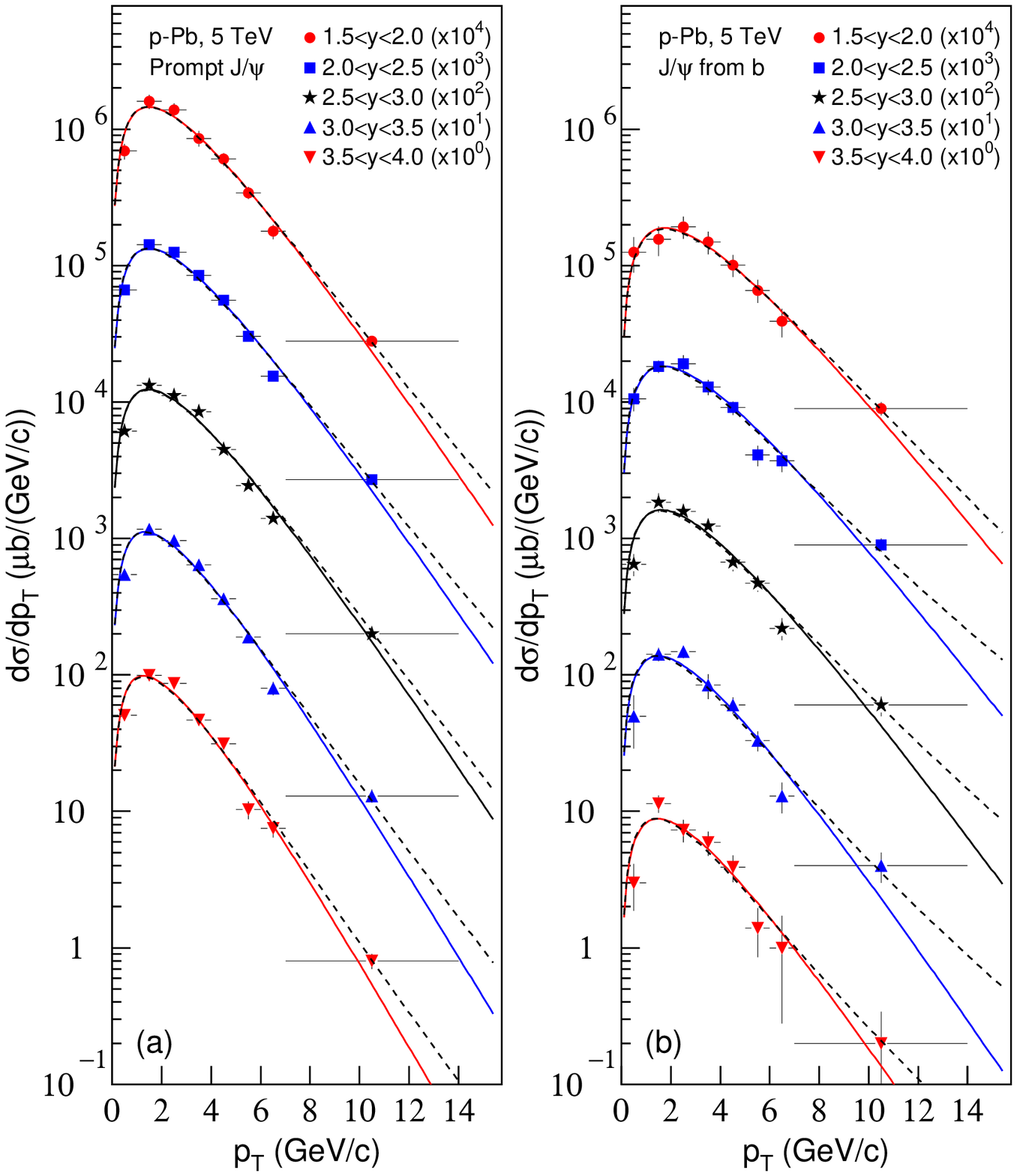}
\end{center}
\vskip0.0cm Fig. 3. Transverse momentum distributions of (a)
prompt $J/\psi$ and (b) $J/\psi$ from $b$ in different rapidity
ranges in $p$-Pb collisions at 5 TeV. The symbols represent the
experimental data of the LHCb Collaboration [16] and the curves
are our calculated results.
\end{figure}

In the above comparisons, each parton in $J/\psi$ production
contributes an exponential distribution with the mean value of
$\langle p_{Ti} \rangle$. The total contribution of two partons is
the folding result of two exponential distributions. This results
in the simplest Erlang distribution with the mean value of
$2\langle p_{Ti} \rangle$ which can be extracted from the
experimental transverse momentum distribution. Assuming an
isotropic emission in the source rest frame, the rapidity and
pseudorapidity distributions are obtained in the multisource
picture. The parameter $\langle p_{Ti} \rangle$ reflects the
violent degree of parton-parton collisions. Because of larger
energy loss in the central region, the parton-parton collisions in
the central region are more violent than those in the
backward/forward regions.

From Figs. 2 and 3 we notice that, although the simplest Erlang
distribution describes the mean trend of experimental $p_T$
distribution in most cases, the theoretical curve seems to
underestimate the tail part of the data, in particular for the
last data. In fact, the simplest Erlang distribution does not
describe the last two data simultaneously. To change this
situation, we revise Eq. (2) to the simplest two-component Erlang
distribution
\begin{equation}
f(p_{T})=\frac{k_{\rm 1st} p_T}{\langle p_{Ti} \rangle_{\rm 1st}
^2} \exp \biggr( -\frac{p_{T}}{\langle p_{Ti} \rangle_{\rm 1st}}
\biggr) + \frac{(1-k_{\rm 1st}) p_T}{\langle p_{Ti} \rangle_{\rm
2nd} ^2} \exp \biggr( -\frac{p_{T}}{\langle p_{Ti} \rangle_{\rm
2nd}} \biggr),
\end{equation}
where $k_{\rm 1st}$ denotes the contribution ratio (relative
contribution) of the first component to the total one. Because
$J/\psi$ is believed to be produced via the hard process. Both the
first and second components should correspond to hard processes.
To give a distinction for the two components, the first component
is regarded as the hard ``peripheral" parton-parton collisions,
and the second one is regarded as the harder ``central"
parton-parton collisions. The mean transverse momentum $\langle
p_T \rangle$ can be given by $2[k_{\rm 1st}\langle p_{Ti}
\rangle_{\rm 1st} +(1-k_{\rm 1st})\langle p_{Ti} \rangle_{\rm
2nd}]$.

By using the simplest two-component Erlang distribution, we
recalculate and show the new $p_T$ distributions by the dashed
curves in Figs. 2 and 3. In the recalculation, we take $k_{\rm
1st}=0.80\pm0.10$ and $\sigma_0$ to be the same as that in Table
1. Other parameters and $\chi^2$/dof are listed in Table 2. One
can see that both $\langle p_{Ti} \rangle_{\rm 1st}$ and $\langle
p_{Ti} \rangle_{\rm 2nd}$ decrease with increase of the rapidity.
The first component determines the peak position and the second
one determines the tail slope. The violent degree of ``peripheral"
parton-parton collisions is lower than that of ``central"
parton-parton collisions.

In our recent work [33], we have used the similar picture and the
same Erlang distribution to describe the transverse momentum
distributions of light particles ($\pi^{\mp}$, $K^{\mp}$, $\bar
p$, and $p$) produced in $p$-Pb collisions at 5 TeV and in Pb-Pb
collisions at 2.76 TeV. The first component corresponds to the
soft excitation process which has 2--5 partons to take part in the
strong interactions. The second component corresponds to the hard
scattering process which has 2 partons to take part in the violent
collision. In the present work, for heavy particles such as
$J/\psi$ mesons and in terms of the two-component, both the
numbers of partons corresponding to the first and second
components are 2, and the first and second components correspond
to hard and harder processes respectively. Although we have used
different explanations for the first and second components in our
previous and present works, they have the same formulism in
principle.

The main goal of the present work is to study some universal laws
existed in high energy collisions. The two-component Erlang
distribution is one of the universal laws. It is known that hard
QCD (Quantum Chromodynamics) contributions follow power law and
not exponential (Erlang) distribution. The fits are very good the
reason may be the two exponential (Erlang) slopes mimic a power
law type behavior. In fact if one has a higher $p_T$ data they may
find a need of third (or more) exponential (Erlang) distribution
with a third slope. Although we interpret the first term in Eq.
(11) as soft and the second term as hard contribution in our
earlier publication [33] which fits light hadrons, the present
work treats heavy particles $J/\psi$ and thus the first term is
interpreted as hard and the second term as harder contribution. In
fact, as one of the universal laws, the two-component Erlang
distribution has more than one interpretations.

Although the production of some $J/\psi$'s can be explained as
thermal recombination of primordially produced $c$ and $\bar c$
quarks at the hadronization transition [53], the present work does
not limit the production process to thermal correlations such as
equilibrium, local equilibrium, non-equilibrium, temperature, and
so forth. Instead, we regard the exponential and Erlang
distributions as statistical laws existed universally in particle
collision and production, nuclear decay and fragmentation, mean
free path, and other topics such as plant seed mass and size [54].
In these topics, many factors affect the results. Each factor
contributes an exponential distribution. The contribution of many
factors is the Erlang distribution which is the folding result of
many exponential distributions. In addition, although the model
used in the present work is called the multisource thermal model,
it may not relate to thermal correlations. In fact, it can also be
a statistical model. Not only for nucleus-nucleus and
proton-nucleus collisions, but also for proton-proton and
electron-positron collisions, the model can be applied in the case
of statistical amount being high.
\\

{\small {Table 2. Values of parameters and $\chi^2$/dof
corresponding to the dashed curves in Figs. 2 and 3. For all of
the cases, $k_{\rm 1st}=0.80\pm0.10$, $\sigma_0$ is the same as
that in Table 1.
{%
\begin{center}
\begin{tabular}{ccccc}
\hline\hline  Fig. & Type or $y$ range & $\langle
p_{Ti}\rangle_{\rm 1st}$ (GeV/$c$) & $\langle p_{Ti}\rangle_{\rm 2nd}$ (GeV/$c$) & $\chi^2$/dof \\
\hline
2(a) & prompt $J/\psi$   & $1.31\pm0.07$ & $1.80\pm0.16$ & 0.917 \\
     & $J/\psi$ from $b$ & $1.45\pm0.09$ & $2.41\pm0.29$ & 0.291 \\
2(b) & prompt $J/\psi$   & $1.16\pm0.04$ & $1.67\pm0.14$ & 1.161 \\
     & $J/\psi$ from $b$ & $1.42\pm0.09$ & $2.10\pm0.22$ & 0.323 \\
2(c) & $2.03<y<3.53$     & $1.33\pm0.07$ & $1.85\pm0.17$ & 0.913 \\
     & $-4.46<y<-2.96$   & $1.18\pm0.05$ & $1.55\pm0.10$ & 0.877 \\
3(a) & $1.5<y<2.0$   & $1.39\pm0.08$ & $1.92\pm0.19$ & 0.258 \\
     & $2.0<y<2.5$   & $1.39\pm0.08$ & $1.96\pm0.20$ & 0.415 \\
     & $2.5<y<3.0$   & $1.39\pm0.08$ & $1.82\pm0.17$ & 0.561 \\
     & $3.0<y<3.5$   & $1.25\pm0.05$ & $1.76\pm0.16$ & 0.682 \\
     & $3.5<y<4.0$   & $1.20\pm0.04$ & $1.69\pm0.14$ & 0.602 \\
3(b) & $1.5<y<2.0$   & $1.65\pm0.13$ & $2.38\pm0.28$ & 0.125 \\
     & $2.0<y<2.5$   & $1.53\pm0.11$ & $2.65\pm0.33$ & 0.281 \\
     & $2.5<y<3.0$   & $1.48\pm0.10$ & $2.50\pm0.30$ & 0.487 \\
     & $3.0<y<3.5$   & $1.34\pm0.07$ & $2.39\pm0.28$ & 0.587 \\
     & $3.5<y<4.0$   & $1.33\pm0.07$ & $2.30\pm0.26$ & 0.532 \\
\hline\hline
\end{tabular}%
\end{center}
}} }

\vskip0.5cm

It is noticed that the two- or multi-component Erlang distribution
has wide applications. In our previous work [55], this
distribution was used to describe multiplicity, mass, transverse
mass, transverse energy, and transverse momentum spectra of
final-state particles in proton-antiproton and electron-proton
(positron-proton) collisions, as well as excitation energy
spectrum for selected events in nucleus-nucleus collisions at high
energies. This distribution was also used to describe the
event-by-event fluctuations in the multiplicity, the total
transverse energy, the mean transverse energy, and the mean
transverse momentum in nucleus-nucleus collisions at high energies
[22], and to describe the production cross-section of
projectile-like isotopes in nucleus-nucleus collisions at
intermediate and high energies [23, 24]. The present work uses
this distribution to a wider range which deals with the hard
process in $J/\psi$ productions at the LHC.
\\

{\section{Conclusions}}

We summarize here our main observations and conclusions.

(a) The rapidity distributions of prompt $J/\psi$, $J/\psi$ from
$b$, and inclusive $J/\psi$ produced in asymmetric $p$-Pb
collisions at 5 TeV can be described by the multisource thermal
model. The sources for the three creations have nearly the same
rapidity shifts in the uncertainty ranges in the backward target
(Pb) and forward projectile ($p$) regions respectively, which
renders that the three sources have the same contributors which
are partons with the same collision energies. The rapidity shift
in the backward Pb-region seems to be greater than that in the
forward $p$-region due to the target having a stronger penetrating
power than the projectile.

(b) The pseudorapidity distributions of prompt $J/\psi$, $J/\psi$
from $b$, and inclusive $J/\psi$ produced in $p$-Pb collisions at
5 TeV are obtained from the parameter values extracted from the
rapidity distributions. The obvious difference between the
pseudorapidity and rapidity distributions is observed due to heavy
particles. In fact, for heavy particles such as $J/\psi$ mesons,
we cannot neglect the difference between the two distributions
even at the LHC energy. It is conceivable that the difference
between the two distributions cannot be neglect at the lower GeV
energy. The best treatment method in the calculation is to
distinguish absolutely the rapidity and pseudorapidity
distributions.

(c) In the considered range, the transverse momentum distributions
of $J/\psi$ mesons can be described by the simplest Erlang
distribution which is the folding result of two exponential
distributions which are contributed by target and projectile
partons respectively. The extracted value of parameter $\langle
p_{Ti} \rangle$ for $J/\psi$ from $b$ is greater than that for
prompt $J/\psi$, and both the values decrease with increase of the
rapidity. The Erlang distribution is an universal law existed in
particle and nuclear physics, even in other fields of nature such
as plant seed. To underline its physics behind is still an open
question.

(d) The parameter $\langle p_{Ti} \rangle$ reflects the violent
degree of parton-parton collisions. Because of larger energy loss
in the central region, the parton-parton collisions in the central
region are more violent than those in the backward/forward
regions. The mean transverse momentum $\langle p_T \rangle$ can be
given by $2\langle p_{Ti} \rangle$ due to the contributions of two
partons. In terms of the two-component, both the numbers of
partons corresponding to the first and second components are 2,
and the first and second components correspond to hard and harder
processes respectively. The mean transverse momentum $\langle p_T
\rangle$ can be given by $2[k_{\rm 1st}\langle p_{Ti} \rangle_{\rm
1st} +(1-k_{\rm 1st})\langle p_{Ti} \rangle_{\rm 2nd}]$.
\\
\\

{\bf Acknowledgment}

This work was supported by the National Natural Science Foundation
of China under Grant No. 11575103 and the US DOE under contract
DE-FG02-87ER40331.A008.

\end{document}